\documentstyle[11pt,epsf]{article}

\def\indspace{\hspace*{1.0em} }
\def\appendix{\setcounter{section}{0}
\def\thesection{Appendix \Alph{section}}
\def\theequation{\Alph{section}.\arabic{equation}}}

\newfont{\subsub}{cmr6}

\newcounter{szk}

\begin{document}
\title{The uniqueness of the profits distribution function \\
in the middle scale region
}
\author{
\footnote{e-mail address: ishikawa@kanazawa-gu.ac.jp} Atushi Ishikawa
\\
Kanazawa Gakuin University, Kanazawa 920-1392, Japan
}
\date{}
\maketitle

\begin{abstract}
\indent
We report the proof that
the expression of extended Gibrat's law is unique and 
the probability distribution function (pdf) is also uniquely derived from
the law of detailed balance and the extended Gibrat's law.
In the proof, two approximations are employed that
the pdf of growth rate is described as tent-shaped exponential functions
and that the value of the origin of growth rate is constant.
These approximations are confirmed in profits data of Japanese companies
2003 and 2004.
The resultant profits pdf fits with the empirical data with high accuracy.
This guarantees the validity of the approximations.
\end{abstract}
\begin{flushleft}
PACS code : 04.60.Nc\\
Keywords : Econophysics; Pareto law; Gibrat law; Detailed balance
\end{flushleft}

\vspace{1cm}
\section{Introduction}
\label{sec-Introduction}
\indspace
In the large scale region of income, profits, assets, sales and etc ($x$),
the cumulative probability distribution function (pdf) $P(> x)$ obeys a power-law
for $x$ which is larger than a certain threshold $x_0$:
\begin{eqnarray}
    P(> x) \propto x^{-\mu}~~~~{\rm for }~~~~x > x_0~.
    \label{Pareto}
\end{eqnarray}
This power-law and the exponent $\mu$ are called 
Pareto's law and Pareto index, respectively~\cite{Pareto}.
The power-law distribution
is well investigated by using various models in econophysics~\cite{MS}.

Recently,
Fujiwara et al.~\cite{FSAKA} find that
Pareto's law 
can be derived kinematically
from the law of detailed balance and Gibrat's law~\cite{Gibrat}
which are also observed in the large scale region $x > x_0$.
In the proof, they assume no model and only use
these two laws in empirical data.

The detailed balance is time-reversal symmetry ($x_1 \leftrightarrow x_2$):
\begin{eqnarray}
    P_{1 2}(x_1, x_2) = P_{1 2}(x_2, x_1)~.
    \label{Detailed balance}
\end{eqnarray}
Here $x_1$ and $x_2$ are  two successive incomes, 
profits, assets, sales, etc. and
$P_{1 2}(x_1, x_2)$ is the joint pdf.
Gibrat's law states that
the conditional pdf of growth rate $Q(R|x_1)$ 
is independent of the initial value $x_1$: 
\begin{eqnarray}
    Q(R|x_1) = Q(R)~.
    \label{Gibrat}
\end{eqnarray}
Here growth rate $R$ is defined as the ratio $R = x_2/x_1$ and
$Q(R|x_1)$ is defined by using the pdf $P(x_1)$
and the joint pdf $P_{1 R}(x_1, R)$ as 
$Q(R|x_1) = P_{1 R}(x_1, R)/P(x_1)$.

In Ref.~\cite{Ishikawa3},
the kinematics is extended to dynamics
by analyzing data on the assessed value of land in Japan.
In the non-equilibrium system we propose an extension of 
the detailed balance (detailed quasi-balance) as follows
\begin{eqnarray}
    P_{1 2}(x_1, x_2) 
    = P_{1 2}( \left( \frac{x_2}{a} \right)^{1/{\theta}}, a~{x_1}^{\theta})~.
    \label{Detailed quasi-balance}
\end{eqnarray}
From Gibrat's law (\ref{Gibrat}) 
and the detailed quasi-balance (\ref{Detailed quasi-balance}),
we derive Pareto's law with annually varying Pareto index.
The parameters $\theta$, $a$ are related to the change of
Pareto index and the relation is confirmed in the empirical data nicely.

These findings are important
for the progress of econophysics.
Above derivations are, however, valid only in the large scale region
where Gibrat's law (\ref{Gibrat}) holds.
It is well known that Pareto's law is not observed 
below the threshold $x_0$~\cite{Gibrat, Badger}.
The reason is thought to be the breakdown of Gibrat's law~\cite{FSAKA}.
The breakdown of Gibrat's law in empirical data
is reported by Stanley's group~\cite{Stanley1}.
Takayasu et al.~\cite{TTOMS} and Aoyama et al.~\cite{Aoyama} also report that
Gibrat's law does not hold in the middle scale region by using data of Japanese companies.

In Ref.~\cite{Ishikawa2}, Gibrat's law is extended in the middle scale region
by employing profits data of Japanese companies in 2002 and 2003.
We approximate the conditional pdf of profits growth rate 
as so-called tent-shaped exponential functions
\begin{eqnarray}
    Q(R|x_1)&=&d(x_1)~R^{-t_{+}(x_1)-1}~~~~~{\rm for}~~R > 1~,
    \label{tent-shaped1}\\
    Q(R|x_1)&=&d(x_1)~R^{+t_{-}(x_1)-1}~~~~~{\rm for}~~R < 1~.
    \label{tent-shaped2}
\end{eqnarray}
By measuring $t_{\pm}$ we have assumed the $x_1$ dependence to be
\begin{eqnarray}
    t_{\pm}(x_1)=t_{\pm}(x_0) \pm \alpha_{\pm}~\ln \frac{x_1}{x_0}~,
    \label{t}
\end{eqnarray}
and have estimated the parameters as~\cite{Ishikawa}
\begin{eqnarray}
    \alpha_{+} \sim \alpha_{-} &\sim& 0~~~~~~~~~{\rm for}~~x_1 > x_0~,
    \label{alphaH}\\
    \alpha_{+} \sim \alpha_{-} &\neq& 0~~~~~~~~~{\rm for}~~x_{{\rm min}} < x_1 < x_0~,
    \label{alphaM}\\
    t_{+}(x_0) - t_{-}(x_0) &\sim& \mu~.
    \label{mu}
\end{eqnarray}
From the detailed balance (\ref{Detailed balance}) and 
extended Gibrat's law (\ref{t}) -- (\ref{mu}),
we have derived the pdf in the large 
and middle scale region uniformly as follows
\begin{eqnarray}
    P(x) = C x^{-\left(\mu+1\right)}~e^{-\alpha \ln^2 \frac{x}{x_0}}
    ~~~~~~~~~{\rm for}~~x > x_{{\rm min}}~,
    \label{HandM}
\end{eqnarray}
where $\alpha = \left(\alpha_{+}+\alpha_{-}\right)/2$.
This is confirmed in the empirical data.

In this study, we prove that the $x_1$ dependence of $t_{\pm}$ 
(\ref{t}) with $\alpha_+ = \alpha_-$ is unique
if the pdf of growth rate 
is approximated by tent-shaped exponential functions 
(\ref{tent-shaped1}), (\ref{tent-shaped2}).
This means, consequently, 
that the pdf in the large and middle scale region (\ref{HandM})
is also unique
if the $x_1$ dependence of $d(x_1)$ is negligible.
We confirm these approximations in profits data of Japanese companies 2003 and 2004
\cite{TSR}
and show that the pdf (\ref{HandM}) fits with empirical data nicely
by the refined data analysis.

\section{Growth rate distributions of profits in the database}
\label{sec-Growth rate distributions of profits in the database}
\begin{figure}[hbp]
 \centerline{\epsfxsize=0.8\textwidth\epsfbox{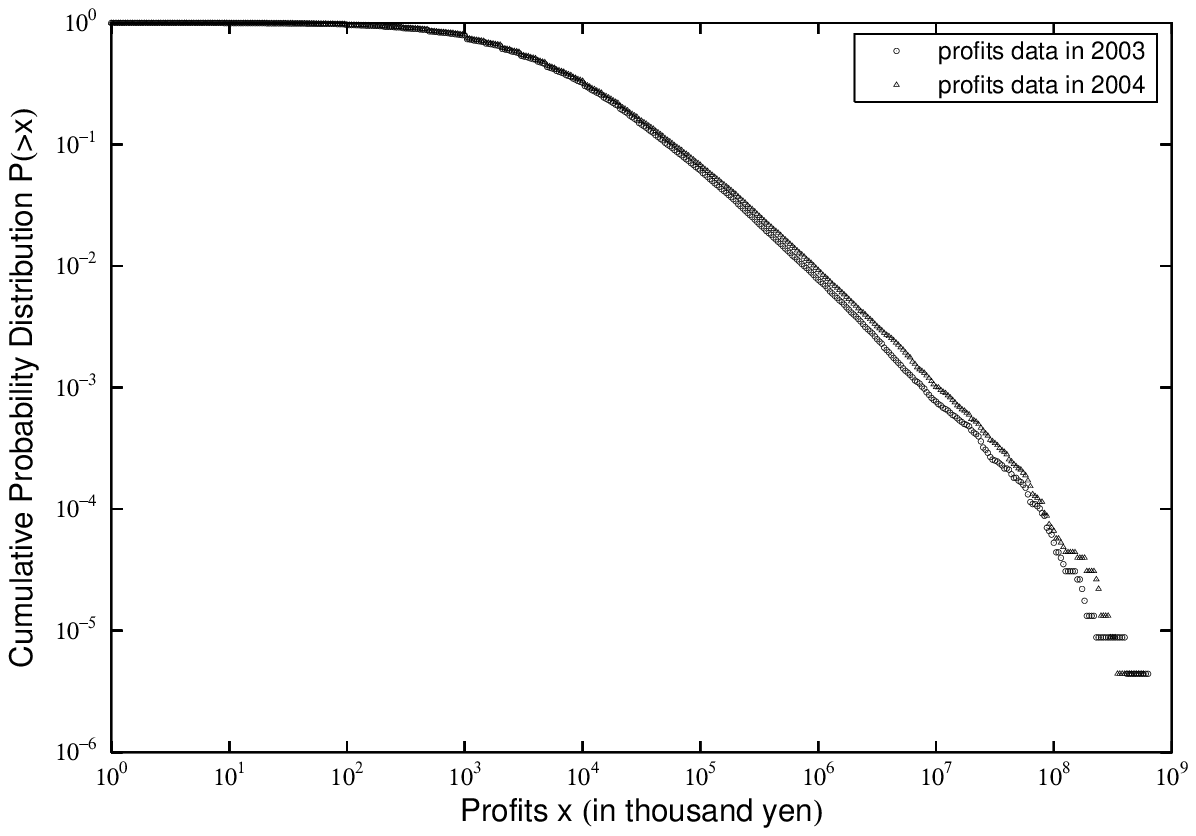}}
 \caption{Cumulative probability distributions $P(> x_1)$ and $P(> x_2)$ for companies, the
 profits of which in 2003 ($x_1$) and 2004 ($x_2$)
 exceeded $0$, $x_1 > 0$ and $x_2 > 0$.
 The number of the companies is ``227,132''.}
 \label{ProfitDistribution}
\end{figure}

\begin{figure}[hbp]
 \centerline{\epsfxsize=0.8\textwidth\epsfbox{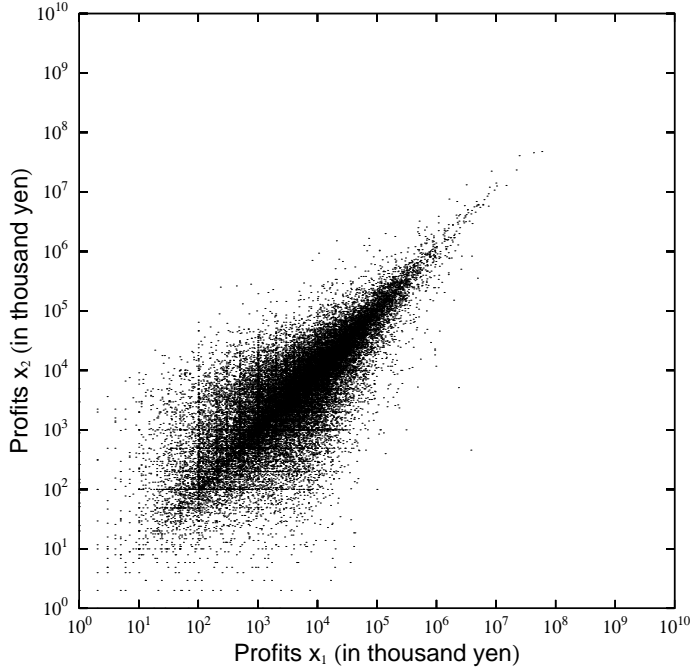}}
 \caption{The scatter plot of all companies in the database, 
 the profits of which in 2003 ($x_1$) and 2004 ($x_2$)
 exceeded $0$, $x_1 > 0$ and $x_2 > 0$.
 The number of the companies is ``227,132''.}
 \label{Profit2003vsProfit2004}
\end{figure}

\indspace
In the database, 
Pareto's law (\ref{Pareto}) is observed in the large scale region
whereas it fails in the middle one (Fig.~\ref{ProfitDistribution}).
At the same time, it is confirmed that the detailed balance (\ref{Detailed balance})
holds not only in the large scale region $x_1 > x_0$ and $x_2 > x_0$
but also in all regions $x_1 > 0$ and $x_2 > 0$ 
(Fig.~\ref{Profit2003vsProfit2004}). \footnote{
The scatter plot in Fig.~\ref{Profit2003vsProfit2004} is different 
from one in Ref~\cite{Ishikawa2}.
The reason is that the identification of profits in 2002 and 2003 in Ref.~\cite{Ishikawa2}
was partly failed.
As a result, the pdfs of profits growth rate 
are slightly different from those in this paper.
The conclusion in Ref.~\cite{Ishikawa2} is, however, not changed. 
}

\begin{figure}[htb]
 \centerline{\epsfxsize=0.8\textwidth\epsfbox{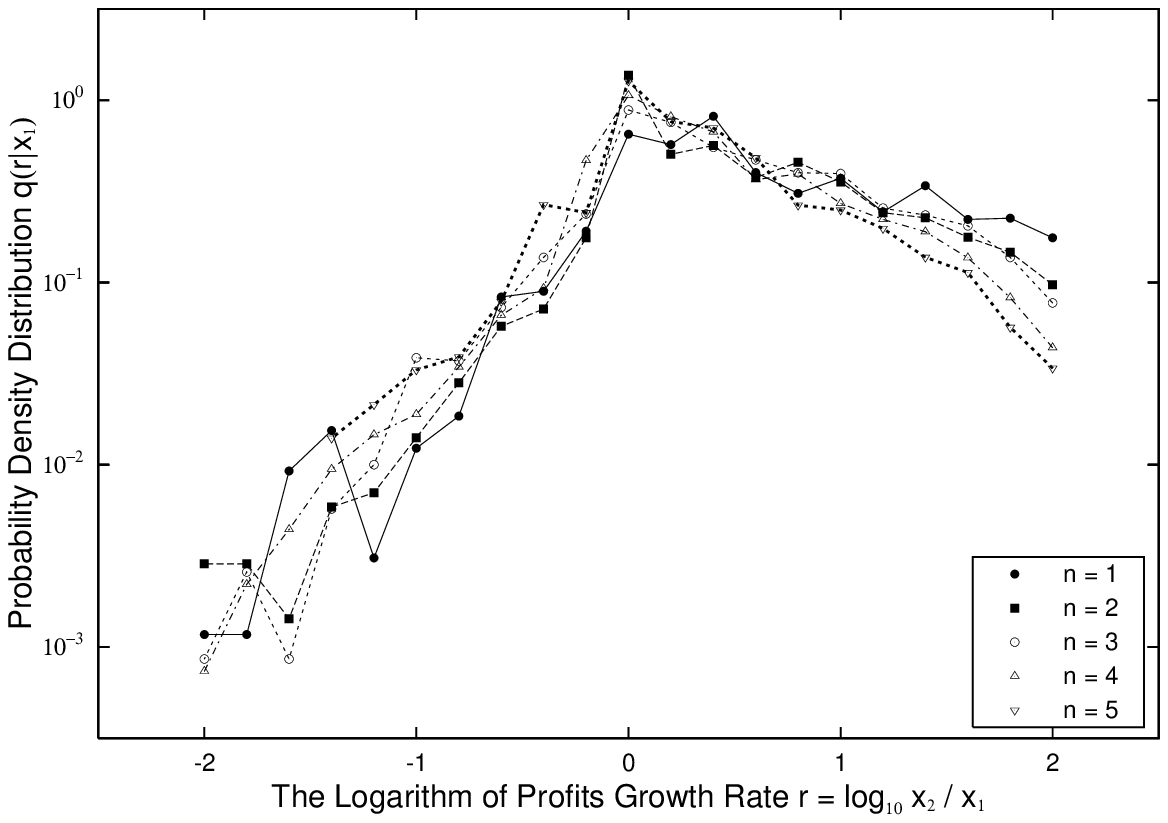}}
 \caption{The probability density distribution $q(r|x_1)$ of the log profits growth rate
 $r = \log_{10} x_2/x_1$ from 2003 to 2004.
 The data points are classified into five 
 bins of the initial profits with equal magnitude in logarithmic scale,
 $x_1 \in 4 \times [10^{1+0.2(n-1)},10^{1+0.2n}]~(n=1, 2, \cdots, 5)$ thousand yen.
 The number of companies in this regime is ``22,005''.}
 \label{ProfitGrowthRateLL}
\end{figure}

\begin{figure}[htb]
 \centerline{\epsfxsize=0.8\textwidth\epsfbox{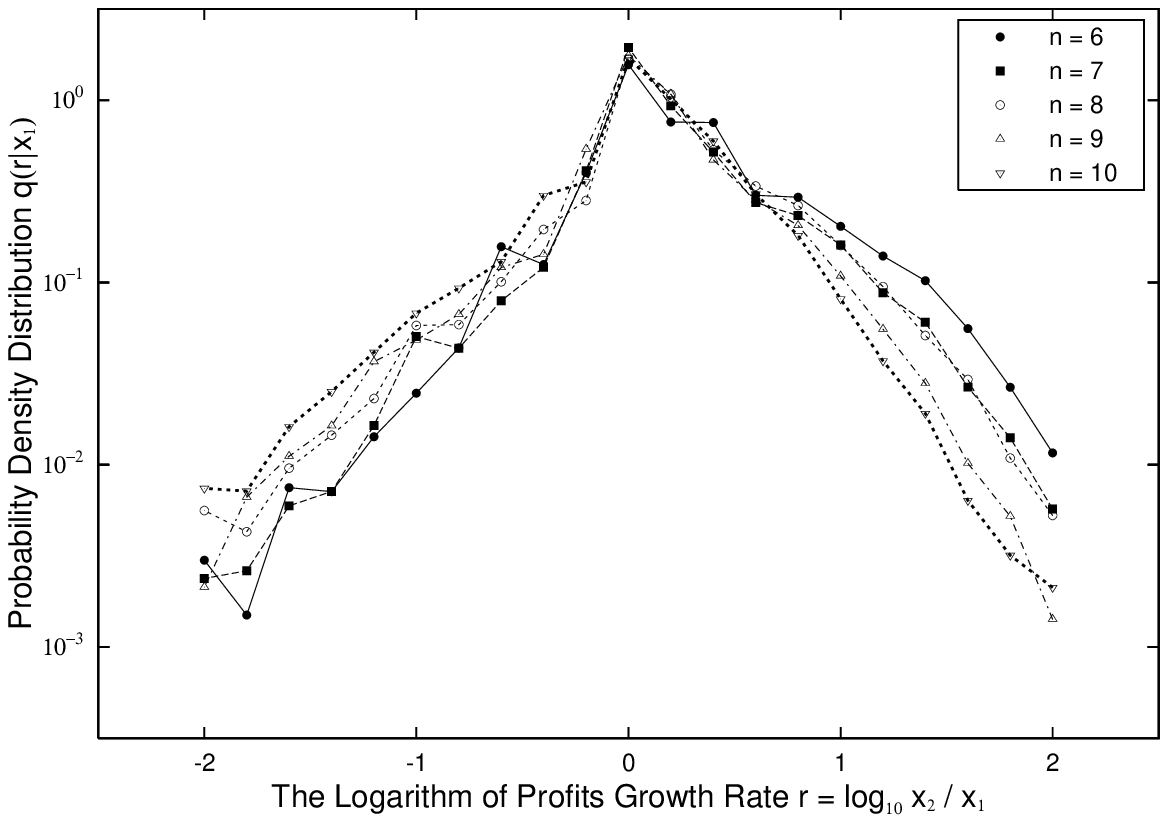}}
 \caption{The probability density distribution $q(r|x_1)$ of the log profits growth rate
 $r = \log_{10} x_2/x_1$ from 2003 to 2004.
 The data points are also classified into five 
 bins of the initial profits with equal magnitude in logarithmic scale,
 $x_1 \in 4 \times [10^{1+0.2(n-1)},10^{2+0.2n}]~(n=6, 7, \cdots, 10)$ thousand yen.
 The number of companies in this regime is ``89,507''.}
 \label{ProfitGrowthRateL}
\end{figure}

\begin{figure}[htb]
 \centerline{\epsfxsize=0.8\textwidth\epsfbox{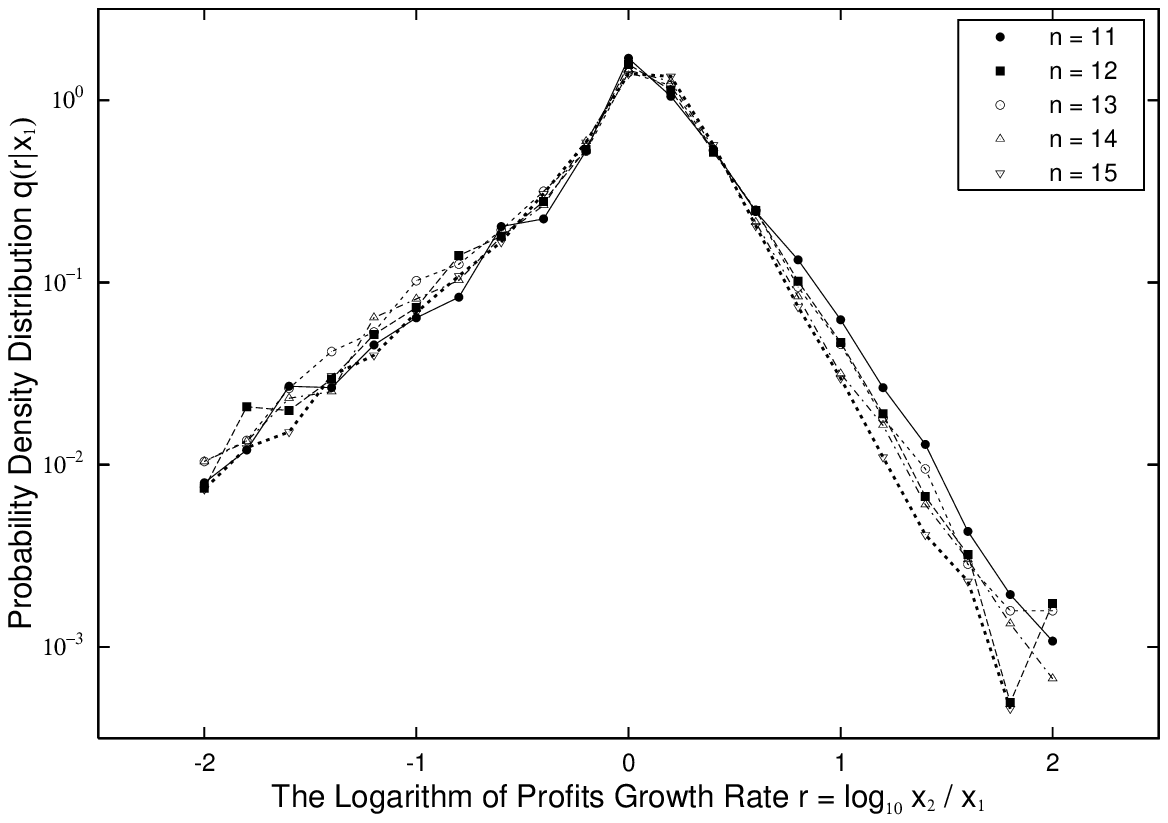}}
 \caption{The probability density distribution $q(r|x_1)$ of the log profits growth rate
 $r= \log_{10} x_2/x_1$ from 2003 to 2004.
 The data points are also classified into five 
 bins of the initial profits with equal magnitude in logarithmic scale,
 $x_1 \in 4 \times [10^{1+0.2(n-1)},10^{1+0.2n}]~(n=11, 12, \cdots, 15)$ thousand yen.
 The number of companies in this regime is ``85,020''.}
 \label{ProfitGrowthRateM}
\end{figure}

\begin{figure}[htb]
 \centerline{\epsfxsize=0.8\textwidth\epsfbox{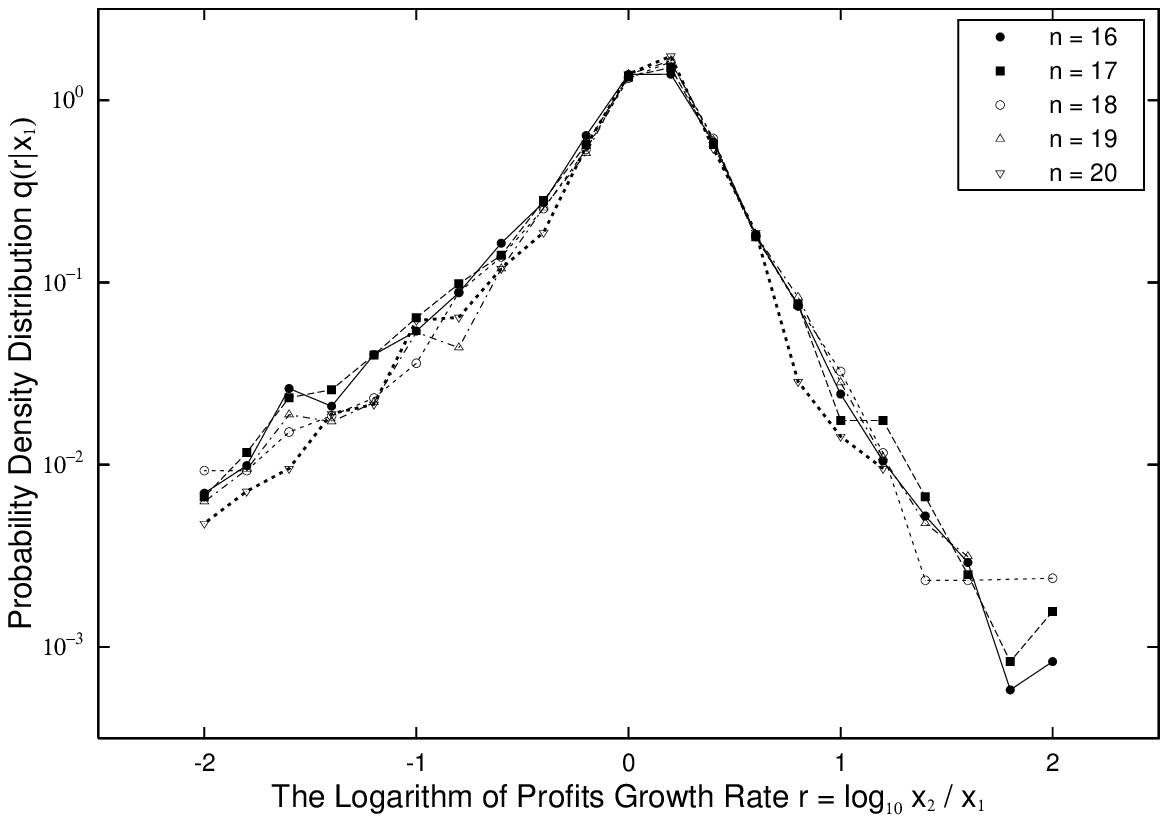}}
 \caption{The probability density distribution $q(r|x_1)$ of the log profits growth rate
 $r= \log_{10} x_2/x_1$ from 2003 to 2004.
 The data points are also classified into five 
 bins of the initial profits with equal magnitude in logarithmic scale,
 $x_1 \in 4 \times [10^{1+0.2(n-1)},10^{1+0.2n}]~(n=16, 17, \cdots, 20)$ thousand yen.
 The number of companies in this regime is ``24,203''.}
 \label{ProfitGrowthRateH}
\end{figure}

The breakdown of Pareto's law is thought to be caused by
the breakdown of Gibrat's law in the middle scale region.
We examine, therefore, the pdf of profits growth rate in the database.
In the analysis, we divide the range of $x_1$ into logarithmically equal bins
as $x_1 \in 4 \times [10^{1+0.2(n-1)},10^{1+0.2n}]$ thousand yen
with $n=1, 2, \cdots, 20$.
In Fig.~\ref{ProfitGrowthRateLL} -- \ref{ProfitGrowthRateH},
the probability densities for $r$ 
are expressed in the case of $n=1, \cdots, 5$,
$n=6, \cdots, 10$, 
$n=11, \cdots, 15$ and $n=16, \cdots, 20$,
respectively.
The number of the companies in Fig.~\ref{ProfitGrowthRateLL} -- \ref{ProfitGrowthRateH}
is ``$22,005$", ``$89,507$", ``$85,020$" and ``$24,203$", respectively.
Here we use the log profits growth rate $r=\log_{10} R$.
The probability density for $r$ defined by $q(r|x_1)$ is
related to that for $R$ by
\begin{eqnarray}
    \log_{10}Q(R|x_1)+r+\log_{10}(\ln 10)=\log_{10}q(r|x_1)~.
\label{Qandq}
\end{eqnarray}

From Fig.~\ref{ProfitGrowthRateLL} -- \ref{ProfitGrowthRateH},
$\log_{10}q(r|x_1)$ is approximated by linear functions of $r$ as follows
\begin{eqnarray}
    \log_{10}q(r|x_1)&=&c(x_1)-t_{+}(x_1)~r~~~~~{\rm for}~~r > 0~,
    \label{approximation1}\\
    \log_{10}q(r|x_1)&=&c(x_1)+t_{-}(x_1)~r~~~~~{\rm for}~~r < 0~.
    \label{approximation2}
\end{eqnarray}
These are expressed 
as tent-shaped exponential functions (\ref{tent-shaped1}), (\ref{tent-shaped2})
by $d(x_1)=10^{c(x_1)}/{\ln 10}$~.
In addition, the $x_1$ dependence of $c(x_1)$ ($d(x_1)$) 
is negligible for $n = 6, \cdots, 20$.
The validity of these approximations should be checked against the results.

\section{The uniqueness of extended Gibrat's law}
\label{sec-The uniqueness of extended Gibrat's law}
\indspace
In this section, we show that
the $x_1$ dependence of $t_{\pm}$ (\ref{t}) is unique
under approximations (\ref{tent-shaped1}), (\ref{tent-shaped2})
((\ref{approximation1}), (\ref{approximation2})).

Due to the relation of
$P_{1 2}(x_1, x_2)dx_1 dx_2 = P_{1 R}(x_1, R)dx_1 dR$
under the change of variables from $(x_1, x_2)$ to $(x_1, R)$,
these two joint pdfs are related to each other
    $P_{1 R}(x_1, R) = x_1 P_{1 2}(x_1, x_2)$.
By the use of this relation, the detailed balance (\ref{Detailed balance})
is rewritten in terms of $P_{1 R}(x_1, R)$ as follows:
\begin{eqnarray}
    P_{1 R}(x_1, R) = R^{-1} P_{1 R}(x_2, R^{-1}).
\end{eqnarray}
Substituting the joint pdf $P_{1 R}(x_1, R)$ for the conditional probability $Q(R|x_1)$,
the detailed balance is expressed as
\begin{eqnarray}
    \frac{P(x_1)}{P(x_2)} = \frac{1}{R} \frac{Q(R^{-1}|x_2)}{Q(R|x_1)}~.
\end{eqnarray}

Under approximations (\ref{tent-shaped1}) and (\ref{tent-shaped2}),
the detailed balance is reduced to
\begin{eqnarray}
    \frac{P(x_1)}{P(x_2)} = \frac{d(x_2)}{d(x_1)}~ R^{+t_{+}(x_1)-t_{-}(x_2)+1}
    \label{}
\end{eqnarray}
for $R>1$.
By using the notation $\tilde{P}(x) \equiv P(x) d(x)$, the detailed balance becomes
\begin{eqnarray}
    \frac{\tilde{P}(x_1)}{\tilde{P}(R~x_1)} = R^{+t_{+}(x_1)-t_{-}(R~x_1)+1}~.
    \label{DE0}
\end{eqnarray}
By expanding Eq.~(\ref{DE0}) around $R=1$, the
following differential equation is obtained 
\begin{eqnarray}
    \Bigl[1+t_{+}(x)-t_{-}(x) \Bigr] \tilde{P}(x) 
        + x~ {\tilde{P}}^{'}(x) = 0,
\end{eqnarray}
where $x$ denotes $x_1$.
The same differential equation is obtained for $R<1$.
The solution is given by
\begin{eqnarray}
    \tilde{P}(x) = C x^{-1}~e^{-G(x)}~,
    \label{HandM3}
\end{eqnarray}
where $t_+(x) - t_-(x) \equiv g(x)$ and $\int g(x)/x~dx \equiv G(x)$.

In order to make the solution (\ref{HandM3}) around $R=1$
satisfies Eq.~(\ref{DE0}),
the following equation must be valid for all $R$:
\begin{eqnarray}
    -G(x)+G(R~x) = \Bigl[t_{+}(x)-t_{-}(R~x) \Bigr] \ln R~.
    \label{Kouho}
\end{eqnarray}
The derivative of Eq.~(\ref{Kouho}) with respect to $x$ is
\begin{eqnarray}
    -\frac{g(x)}{x}+\frac{g(R~x)}{x} 
        = \Bigl[{t_{+}}^{'}(x)-R~{t_{-}}^{'}(R~x) \Bigr] \ln R~.
        \label{Kouho2}
\end{eqnarray}
By expanding Eq.~(\ref{Kouho2}) around $R=1$, 
following differential equations are obtained
\begin{eqnarray}
    &&x~\Bigl[{t_{+}}^{''}(x)+{t_{-}}^{''}(x) \Bigr]
        +{t_{+}}^{'}(x)+{t_{-}}^{'}(x)=0~,\\
    &&2~{t_{+}}^{'}(x)+{t_{-}}^{'}(x)-3x~{t_{-}}^{''}(x)
        -x^2~\Bigl[{t_{+}}^{(3)}(x)+2~{t_{-}}^{(3)}(x) \Bigr]=0~.        
\end{eqnarray}
The solutions are given by
\begin{eqnarray}
    t_+(x) &=& -\frac{C_{-2}}{2} \ln^2 x 
        + \left(C_{+1}-C_{-1} \right) \ln x + \left( C_{+0}-C_{-0} \right)~,\\
    t_-(x) &=& \frac{C_{-2}}{2} \ln^2 x + C_{-1} \ln x + C_{-0}~.
\end{eqnarray}
To make these solutions satisfy Eq.~(\ref{Kouho}), the coefficients must be
$C_{-2}=0$ and $C_{+1}=0$.
Finally we conclude that $t_{\pm}(x)$ is uniquely expressed as 
Eq.~(\ref{t}) with $\alpha_+ = \alpha_-$.
\section{The profits distribution and the data fitting}
\label{sec-The profits distribution and the data fitting}
\indspace
Under approximations (\ref{tent-shaped1}) and (\ref{tent-shaped2})
((\ref{approximation1}) and (\ref{approximation2})),
we obtain the profits pdf
\begin{eqnarray}
    \tilde{P}(x) = P(x) d(x) 
    = C x^{-\left(\mu+1\right)}~e^{-\alpha \ln^2 \frac{x}{x_0}}~,
    \label{HandM2}
\end{eqnarray}
where we use the relation (\ref{mu}) confirmed in Ref.~\cite{Ishikawa}.

\begin{figure}[htb]
 \centerline{\epsfxsize=0.8\textwidth\epsfbox{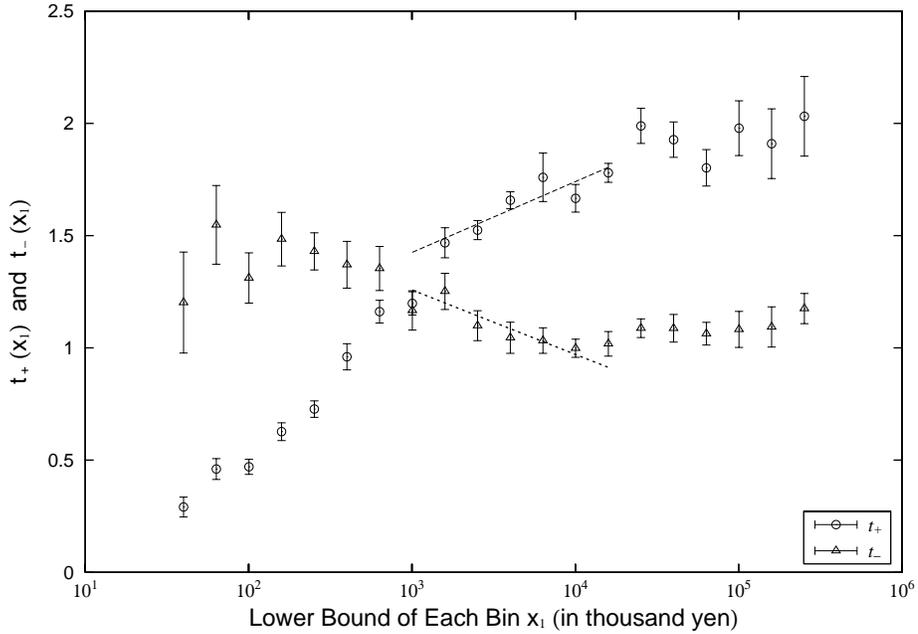}}
 \caption{
 The relation between the lower bound of each bin $x_1$ and $t_{\pm}(x_1)$.
 From the left, each data point represents $n=1, 2, \cdots, 20$.
 }
 \label{X1vsT}
\end{figure}

\begin{figure}[htb]
 \centerline{\epsfxsize=0.8\textwidth\epsfbox{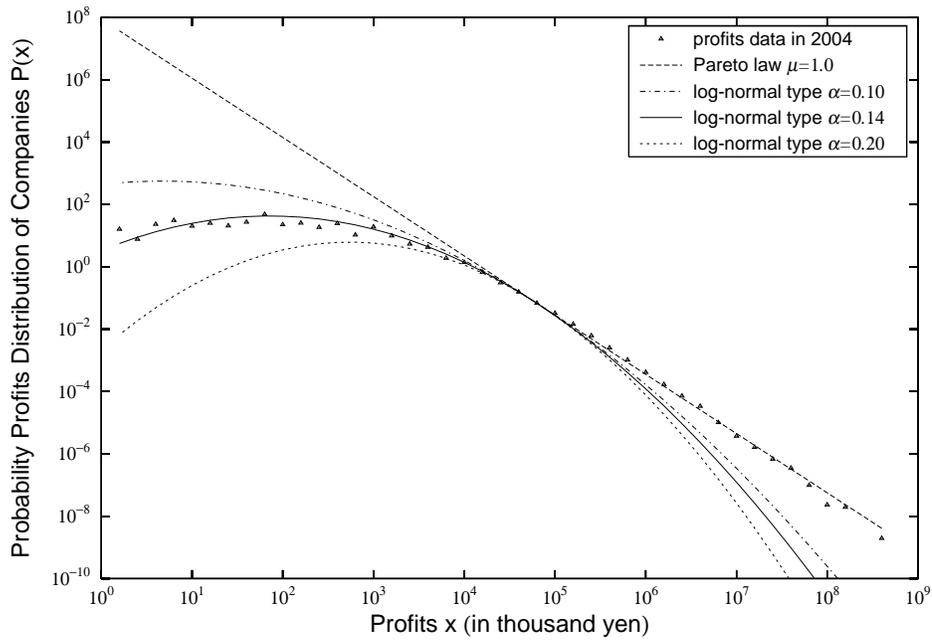}}
 \caption{
 The probability distribution function (pdf) $P(x_2)$ for companies, the
 profits of which in 2003 ($x_1$) and 2004 ($x_2$)
 exceeded $0$, $x_1 > 0$ and $x_2 > 0$.
 The pdf derived from the detailed balance
 and the extended Gibrat's law fits with the data accurately.
  Indices $\mu$, $\alpha$ and $x_0$ are already
  estimated in the extended Gibrat's law.
  If the parameters are different from the estimation,
  the pdf cannot fit with the data
  ($\alpha=0.10$ or $\alpha=0.20$ for instance).
 }
 \label{ProfitDistributionFit}
\end{figure}

In Fig.~\ref{X1vsT}, 
$t_{\pm}$ hardly responds to 
$x_1$ for $n=17, \cdots, 20$.
This means that Gibrat's law holds in the large profits region.
On the other hand,
$t_{+}$ linearly increases and $t_{-}$ linearly decreases 
symmetrically with $\log_{10} x_1$
for $n=9, 10, \cdots, 13$.
The parameters are estimated 
as Eq.~(\ref{alphaH}) and (\ref{alphaM}) with
$\alpha$ ($= \alpha_+ = \alpha_-) \sim 0.14$
and $x_0 = 4 \times 10^{1+0.2(17-1)} \sim 63,000$ thousand yen.
Because the $x_1$ dependence of $c(x_1)$ ($d(x_1)$) 
is negligible in this region, 
the profits pdf is reduced to Eq.~(\ref{HandM}).
We observe that this pdf fits with the empirical data nicely 
in Fig.~\ref{ProfitDistributionFit}.

Notice that the estimation of $\alpha$ in Fig.~\ref{X1vsT} is significant.
If we take a slightly different $\alpha$, the pdf (\ref{HandM}) 
cannot fit with the empirical data
($\alpha = 0.10$ or $\alpha = 0.20$ 
in Fig.~\ref{ProfitDistributionFit} for instance).
\section{Conclusion}
\label{sec-Conclusion}
\indspace
In this paper,
we have shown the proof that
the expression of extended Gibrat's law is unique and 
the pdf in the large and middle scale region is also uniquely derived from
the law of detailed balance and the extended Gibrat's law.
In the proof, we have employed two approximations that
the pdf of growth rate is described as tent-shaped exponential functions
and that the value of the origin of growth rate is constant.
These approximations have been confirmed in profits data of Japanese companies
2003 and 2004.
The resultant pdf of profits has fitted with the empirical data with high accuracy.
This guarantees the validity of the approximations.

For profits data we have used, the distribution is power in the large scale region
and log-normal type in the middle one.
This does not claim that
all the distributions in the middle scale region are log-normal types.
For instance, the pdf of personal income growth rate or sales of company
is different from tent-shaped exponential functions~\cite{FSAKA}.
In this case, the extended Gibrat's law takes a different form.
In addition, we describe no pdf in the small scale region~\cite{Yakovenko}.
Because the $x_1$ dependence of $d(x_1)$ in this region is not negligible
(Fig.~\ref{ProfitGrowthRateLL}).

Against these restrictions, the proof and the method in this paper
is significant for the investigation of distributions in the middle and small scale region.
We will report the study about these issues in the near future.


\section*{Acknowledgments}
\indent

The author is 
grateful to 
the Yukawa Institute for Theoretical 
Physics at Kyoto University,
where this work was initiated during the YITP-W-05-07 on
``Econophysics II -- Physics-based approach to Economic and
Social phenomena --'',
and especially to 
Professor~H. Aoyama for the critical question
about the author's presentation.
Thanks are also due to  Dr.~Y. Fujiwara
for a lot of useful discussions and comments.





\end{document}